\documentclass[preprint,aps]{revtex4}
\usepackage{epsf}
\usepackage{graphicx} 
\usepackage{color}

\begin{document}

\title
{High-Order Coupled Cluster Method Study of Frustrated and Unfrustrated 
Quantum Magnets in External Magnetic Fields}
\author
{
D. J. J. Farnell$^1$, R. Zinke$^2$, J. Richter$^2$, and J. Schulenburg$^3$
}
\affiliation{$^1$Academic Department of Radiation Oncology, Division of Cancer Studies, Faculty of Medical and Human Science,  University of Manchester, c/o Christie Hospital NHS Foundation Trust, Manchester M20 4BX, United Kingdom}
\affiliation{$^2$Institut f\"ur Theoretische Physik, Otto-von-Guericke Universit\"at
Magdeburg, P.O.B. 4120, 39016 Magdeburg, Germany}
\affiliation{$^3$Universita\"atsrechzenzentrum, Otto-von-Guericke Universit\"at
Magdeburg, P.O.B. 4120, 39016 Magdeburg, Germany}

\date{\today}

\begin{abstract}
We apply the coupled cluster method (CCM) in order to study the 
ground-state properties of  the (unfrustrated) square-lattice and (frustrated) 
triangular-lattice spin-half Heisenberg antiferromagnets in the presence of 
external magnetic fields. Approximate methods are difficult to apply to the 
triangular-lattice antiferromagnet  because of frustration, and so, for example, 
the quantum Monte Carlo (QMC) method suffers from the ``sign problem.'' 
Results for this model in the presence of magnetic field are rarer than 
those for the square-lattice system. Here we determine and solve the 
basic CCM equations by using the localised approximation scheme 
commonly referred to as the `LSUB$m$' approximation scheme 
and we carry out high-order calculations by using intensive
computational methods. We calculate the ground-state energy, the uniform 
susceptibility, the total (lattice) magnetisation and the local (sublattice) 
magnetisations as a function of the magnetic field strength. Our results for 
the lattice magnetisation of the square-lattice case compare well
to those results of QMC for all values of the applied external
magnetic field. We find a value for magnetic susceptibility 
of $\chi=0.070$ for the square-lattice antiferromagnet, which is 
also in agreement with the results of other approximate methods 
(e.g., $\chi=0.0669$ via QMC). Our estimate for the range of the extent 
of the ($M/M_s=$)$\frac 13$ magnetisation plateau for the triangular-lattice 
antiferromagnet is $1.37< \lambda < 2.15$, which is in 
good agreement with results of spin-wave theory ($1.248 < \lambda 
<  2.145$) and exact diagonalisations ($1.38 < \lambda < 2.16$). 
Our results therefore support those of exact diagonalisations
that indicate that the plateau begins at a higher value
of $\lambda$ than that suggested by spin-wave theory. 
The CCM value for the in-plane magnetic susceptibility per site 
is $\chi=0.065$, which is below the result of the spin-wave theory 
(evaluated to order $1/S$) of $\chi_{SWT}=0.0794$.
Higher order calculations are thus suggested for both SWT
and CCM LSUB$m$ calculations in order to determine
the value of $\chi$ for the triangular lattice conclusively.
\end{abstract}

\maketitle

\pagebreak

\section{Introduction}

Low-dimensional quantum magnets provide a difficult challenge to the
theoretical physicist because of their strong quantum fluctuations and 
their complex dynamics \cite{diep04,Scholl:2004}. These effects lead to 
rich physics that include novel quantum phases, as well as quantum 
phase transitions between semi-classical magnetically ordered phases
and magnetically disordered quantum phases, see, e.g., Ref.~\cite{Sachdev99}.

An interesting field of research is that of the behaviour of quantum 
magnetic systems in the presence of external magnetic fields, see, e.g.
Refs.~\cite{Hon1999,CGHP,LhuiMi,squareTriangleED,HSR04}. This topic has become more important
by the discovery of exotic parts of the magnetisation curve of quantum
antiferromagnets, such as plateaux and jumps 
\cite{Hon1999,HSR04,nishi,chub,alicea,oshi,SchuRi,jump,ono,squareTriangleED,kagome_pl,schnalle,schroeder,fortune} 
in the lattice magnetisation with respect to the externally applied field. 
Indeed, the presence of these plateaux and jumps may sometimes be linked purely 
to quantum effects because they are not observed in equivalent 
classical models at $T=0$ \cite{jump,kawamura,zhito,cabra}. Clearly, the behaviour of 
quantum magnetic materials in the presence of external magnetic fields 
is an important aspect in their subsequent technological exploitation.
Several methods such as quantum Monte Carlo method (QMC), field theories, 
exact diagonalisation of finite systems, spin-wave techniques and 
strong-coupling approximation have been used
\cite{Hon1999,CGHP,LhuiMi,squareTriangleED,HSR04}
to study these systems. However, each method has its own specific
limitations; for instance, the QMC is restricted (essentially) to 
unfrustrated systems because of the infamous `sign problem.'

In this article we focus on the behaviour of quantum antiferromagnets 
as they react to externally imposed magnetic fields by a method of
quantum many-body theory called the coupled cluster method (CCM)
\cite{ccm1,ccm2,ccm5,ccm12,ccm15,ccm20,ccm26,ccm27,ccm32,ccm35}. 
The CCM has been used previously in order to treat a wide range 
of strongly interacting quantum systems. In particular, the CCM is not 
restricted, in principle, by the spatial dimensionality of the
problem or by the presence of competition between bonds, i.e., in 
frustrated quantum spin systems. A remarkable advance in the accuracy 
of the method for a localised approximation scheme called the LSUB$m$ 
scheme has been afforded by the use of ``high-order'' CCM via
computer-algebraic implementations \cite{ccm12,ccm15,ccm20,ccm26}. 
This computer code developed by DJJ Farnell and J Schulenburg \cite{code} 
is
very flexible in terms of the range of underlying 
crystallographic lattice, spin quantum number, and types
of Hamiltonian that may be studied. Furthermore, recent 
advances to this code now allow ``generalised expectation
values'' (with respect to one-spin and two-spin operators) 
and (separately) excited-state properties to be evaluated 
to high orders of approximation. Indeed, we employ
the new code for the generalised expectation values
to determine the lattice magnetisation and individual 
sublattice magnetisations of quantum antiferromagnets
in external magnetic fields.

The relevant Hamiltonian for an antiferromagnet in an 
external field is defined by 
\begin{equation}
H = \sum_{\langle i,j\rangle} {\bf s}_i ~ \cdot ~ {\bf s}_j - \lambda \sum_i s_i^z
~~ ,
\label{heisenberg}
\end{equation}
where the index $i$ runs over all lattice sites on the lattice. The
expression $\langle i,j\rangle$ indicates a sum over all nearest-neighbour 
pairs, although each pair is counted once and once only. The strength 
of the applied external magnetic field is given by $\lambda$.

The quantum ground states at $\lambda=0$ of all of the cases considered 
here are semi-classically ordered (albeit the classical order is reduced by quantum fluctuations)
\cite{Scholl:2004}. 
Classically, nearest-neighbours align in antiparallel directions for the 
bipartite antiferromagnets such as the antiferromagnet on the square lattice 
and at angles of 120$^{\circ}$ to each other for the Heisenberg
antiferromagnet on the (tripartite) triangular lattice at $\lambda=0$. 
In the presence of an externally applied magnetic field ($\lambda>0$), 
the classical picture indicates that the spins will cant at various 
angles and that at a ``saturation" value of $\lambda=\lambda_s$ 
(square: $\lambda_s=4$; triangle; $\lambda_s=4.5$) all spins align with 
the field. The magnetisation saturates to a maximum value $M=M_s$ 
at this point. 

However, 
we remark that the behaviour of quantum spin-half square-lattice antiferromagnet
in a magnetic field   
\cite{Hon1999,squareTriangleED,square1,runge,zheng,lswt,hamer,luscherlauchli,zhito09}
is (essentially) the same as that of the classical model, albeit modified 
by quantum fluctuations. Second-order (and third-order) spin-wave theory 
\cite{square1,zheng,hamer} thus provides a good approximation to the behaviour 
of this model. Exact diagonalisations and QMC simulations
\cite{squareTriangleED,luscherlauchli} 
also provide good results for this case.  Very recently, in Refs.
\cite{luscherlauchli,zhito09}, 
the field dependence of the low-energy descriptors of this model (i.e., spin stiffness, 
spin-wave velocity, and magnetic susceptibility) 
have been investigated
using exact diagonalisations and spin-wave theory. 
An excellent review of the properties of the spin-half square-lattice antiferromagnet
is given by Ref. \cite{manousakis}. 

By contrast, the behaviour of 
the quantum case for spin-half triangular-lattice antiferromagnet 
\cite{Hon1999,HSR04,nishi,chub,alicea,chub94,trumper00,ono,squareTriangleED} is much different to 
that of the classical model. In particular, a magnetisation plateau is observed 
at $M/M_s=\frac 13$ over a finite region of $\lambda$. The range of
this plateau has been estimated by spin-wave theory \cite{chub,alicea} to 
be given by $1.248<\lambda<2.145$, whereas exact diagonalisations 
\cite{Hon1999,HSR04,nishi,squareTriangleED} predict a region given by $1.38<\lambda<2.16$. 
We note that the application of the QMC method (leading to precise results for
bipartite lattices)  to the case of the triangular  
is severely limited by the ``sign problem" due to frustration. The
available spin-wave and exact-diagonalization data for the triangular lattice
seem to be less accurate and complementary
results are desirable. 
%The existence of a $\frac 13$-plateau is also related to a gap in the 
%excitation spectrum for magnetic fields within the plateau region,
%although this is not considered here. 
Furthermore, recent experimental
evidence \cite{fortune} for the magnetic material Cs$_2$CuBr$_4$ 
suggests that a series of plateaux might exist at values of $M/M_s$ equal to
1/3, 1/2, 5/9 and 2/3. The authors of this article suggest 
that this might be due to unit cells of differing size for the different 
plateaux, e.g., each having an overall magnetisation of 1/2, and furthermore 
that theory has thus far only predicted the first of these at 1/3. However, 
the treatment of these possible higher plateau is beyond the 
scope of this article. 

The main goal of our paper is to explain how the CCM can be used to
investigate the magnetisation process of quantum antiferromagnets and to 
provide detailed CCM results for the spin-half Heisenberg antiferromagnets on the     
square and the triangular lattices.
The CCM has previously been applied with much success to the subject of 
quantum magnetic systems at zero temperature. The CCM provides 
accurate results even in the presence of very strong frustration. 
In particular, the use of computer-algebraic implementations
\cite{ccm12,ccm15,ccm20,ccm26} of the CCM for 
quantum systems of infinite numbers of particles has been 
found to be very effective with respect to these spin-lattice problems.
Here we present a brief description of the CCM  formalism and its 
application via computational methods to the subject of quantum spin models. 
We then describe the application of the method to the spin-half Heisenberg model 
for the square and triangular lattices at zero temperature in the 
presence of an external magnetic field. We present our 
results and then discuss the conclusions of this research.

\section{The Coupled Cluster Method (CCM)}

As the CCM has been discussed extensively elsewhere (see Refs. 
\cite{ccm1,ccm2,ccm5,ccm12,ccm15,ccm20,ccm26,ccm27,ccm32,ccm35}),  
we do not consider the methodology in depth here. In particular, 
the interested reader should note that the use of computer-algebraic 
implementations has been considered in Refs. \cite{ccm12,ccm15,ccm20,ccm26}. 
However, it is still important to remark here that the exact ket 
and bra ground-state energy eigenvectors, $|\Psi\rangle$ and 
$\langle\tilde{\Psi}|$, of a general many-body system described 
by a Hamiltonian $H$, are given by
%%%%%%%%%%%%%%%%%%%%%%%%%%%%%%%%%%
\begin{equation} 
H |\Psi\rangle = E_g |\Psi\rangle
\;; 
\;\;\;  
\langle\tilde{\Psi}| H = E_g \langle\tilde{\Psi}| 
\;. 
\label{eq1} 
\end{equation} 
%%%%%%%%%%%%%%%%%%%%%%%%%%%%%%%%%%  
The ket and bra states are parametrised within the 
%single-reference 
CCM as follows:   
%%%%%%%%%%%%%%%%%% 
\begin{eqnarray} 
|\Psi\rangle = {\rm e}^S |\Phi\rangle \; &;&  
\;\;\; S=\sum_{I \neq 0} {\cal S}_I C_I^{+}  \nonumber \; , \\ 
\langle\tilde{\Psi}| = \langle\Phi| \tilde{S} {\rm e}^{-S} \; &;& 
\;\;\; \tilde{S} =1 + \sum_{I \neq 0} \tilde{{\cal S}}_I C_I^{-} \; .  
\label{eq2} 
\end{eqnarray} 
%%%%%%%%%%%%%%%%%% 
One of the most important features of the CCM is that one uses a 
single model or reference state $|\Phi\rangle$ that is normalised.
This, in turn, leads to a normalisation condition for the 
ground-state bra and ket wave functions ($\langle \tilde\Psi|\Psi\rangle 
\equiv \langle\Phi|\Phi\rangle=1$). The model state is required to have the 
property of being a cyclic vector with respect to two well-defined Abelian 
subalgebras of {\it multi-configurational} creation operators $\{C_I^{+}\}$ 
and their Hermitian-adjoint destruction counterparts $\{ C_I^{-} \equiv 
(C_I^{+})^\dagger \}$. For spin systems the model state $|\Phi\rangle$ 
typically can be chosen as an independent-spin product state 
and the corresponding 
operators $\{C_I^{+}\}$ can be expressed as a product of a set of spin lowering
operators, see below and for more details also Refs. \cite{ccm12,ccm15,ccm20,ccm26}.

\begin{figure}
\epsfxsize=11cm
\centerline{\epsffile{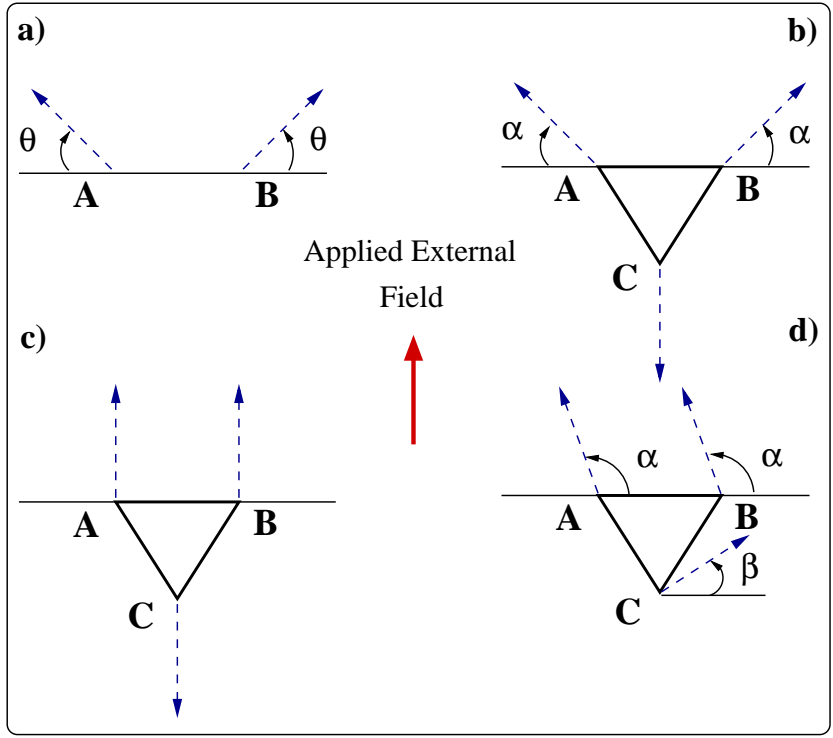}}
\caption{The model states used in the CCM calculations for the Heisenberg model 
in an external magnetic field. a) The bipartite Lattices. Spins on the $A$ and $B$ 
sublattices make angles $\theta$ to the $x$-axis. b) The first model state for 
the triangular lattice (model state I). Spins on the $A$ and $B$ sublattices 
make angles $\alpha$ to the $x$-axis. Spins on the $C$ sublattice point downwards.
c) The second model state for the triangular lattice (model state II).  Spins on 
the $A$ and $B$ sublattices point upwards. Spins on the $C$ sublattice point 
downwards.  d) The third model state for the triangular lattice (model state 
III). Spins on the $A$ and $B$ sublattices make  angles $\alpha$ to the $x$-axis. 
Spins on the $C$  make an angle $\beta$ to the $x$-axis. (Model state II is 
also a limiting case of model states I and III.)}
\label{model_states}
\end{figure}

The CCM formalism is exact in the limit of inclusion of
all possible multi-spin cluster correlations within 
$S$ and $\tilde S$, although this is usually impossible to achieve
practically. It is therefore 
necessary to utilise various approximation schemes 
within $S$ and $\tilde{S}$. Here we use the localised 
LSUB$m$ scheme, in which all 
multi-spin correlations over distinct locales on the 
lattice defined by $m$ or fewer contiguous sites are 
retained.
This approximation scheme has been successfully applied to determine the
ground-state phases of quantum spin systems, see e.g. 
\cite{ccm15,ccm26}.
%We also make the specific restriction that the 
%creation operators $\{C_I^+\}$ in $S$ preserve
%any additional symmetries of the Hamiltonian.
%We denote as distinct configurations those clusters 
%(in an appropriately defined subspace) that are 
%inequivalent under the point- and space-group symmetries 
%of both the lattice and the Hamiltonian. 
%Another important feature 
%of the method is that the bra and ket states are not always 
%explicitly constrained to be Hermitian conjugates when we 
%make such approximations, although the important 
%Helmann-Feynman theorem is always preserved. We remark 
%that the CCM provides results in the infinite-lattice limit 
%$N \rightarrow \infty$ from the outset.
The CCM is a bi-variational formulation in which the 
bra and ket states are parametrised separately. This means that the ket 
and bra states are not explicitly constrained to be Hermitian conjugates. However, an 
advantage of this approach is that the Goldstone linked-cluster 
theorem is obeyed and so results may be found in the infinite-lattice 
limit $N \rightarrow \infty$  from the outset. The important Helmann-Feyman theorem 
is also obeyed at all levels of approximation. The ket-state and bra-state equations 
are obtained using the following formulae,
\begin{eqnarray} 
\langle\Phi|C_I^{-} {\rm e}^{-S} H {\rm e}^S|\Phi\rangle &=& 0 ,  \;\; 
\forall I \neq 0 \;\; ; \label{ket_state_eqn} \\ 
\langle\Phi|\tilde{S} {\rm e}^{-S} [H,C_I^{+}] {\rm e}^S|\Phi\rangle 
&=& 0 , \;\; \forall I \neq 0 \;\; . \label{bra_state_eqn}
\end{eqnarray} 
The method in which Eqs. (\ref{ket_state_eqn}) 
and (\ref{bra_state_eqn}) are solved has been discussed extensively 
elsewhere \cite{ccm1,ccm2,ccm5,ccm12,ccm15,ccm20,ccm26,ccm27,ccm32,ccm35}. 
However, we remark here that the computational method
for solution of the CCM problem may be broken into three parts. 
The first task is, namely, to enumerate the fundamental set of CCM clusters 
for a given level of approximation. Secondly, we must determine the 
ket-state equations in terms of the CCM ket-state correlation 
coefficients by pattern-matching those clusters $C_I^-$ in the 
fundamental set to term in ${\rm e}^{-S} H {\rm e}^S$. Once we 
have determined the ket-state equations, the bra-state equations
may be determined directly. Finally, we solve the coupled CCM 
equations for the ket- and bra-state correlation coefficients, 
e.g., by using the Newton-Raphson method for the ket-state equations.
Expectation values such as the lattice magnetisation may be obtained 
after we have solved for both the ket and bra states. Again, we 
refer the interested reader to Refs. \cite{ccm12,ccm15,ccm20,ccm26} 
for more details of the practicalities of carrying out CCM calculations 
to high order.

Here we use the classical ground states of these systems of 
the Heisenberg model in an external magnetic field as the model state. 
However, the magnitude of the characteristic canting angles in the quantum model
(i.e., the angle between the local directions of the spins and the external magnetic 
field) may be different from the corresponding classical value. 
Hence, we do not choose the classical result for those angles. Indeed, 
we consider the angles as a free parameters in the CCM calculation, 
which has to be determined by minimisation of the CCM ground-state
energy. 
%\cite{ccm6,ccm17,ccm21,ccm25,ccm34}. 

The ground state of the classical system at zero external field 
($\lambda=0$) has nearest neighbouring spins aligning in opposite 
directions for the bipartite lattices (e.g., the square lattice) 
and at angles of 120$^{\circ}$ to each other for the triangular lattice. 
Classically, the spins react to an external magnetic field by 
changing their alignment to that of the direction of the field. 
This is shown in Fig.~\ref{model_states}. For the bipartite lattices, 
the spins thus cant at an angle of $\theta$ and $\pi - \theta$ to 
the $x$-axis, as is shown in Fig.~\ref{model_states}a. 
By contrast, for the tripartite triangular lattice and related frustrated
lattices one ought to distinguish
between an applied field within the plane defined by the 120$^{\circ}$
planar state 
and a field perpendicular to this plane. Although on the
classical level both cases are energetically
equivalent\cite{kawamura,chub,zhito,cabra}, thermal or quantum fluctuations
favour the planar configuration
\cite{kawamura,chub,zhito,cabra}.  Therefore in the present paper we
restrict our considerations to 
planar states and a corresponding magnetic field applied within this plane.
Following Ref.~\cite{chub,ono}
we employ three different model states for the tripartite triangular 
lattice. The first such model state is one in which 
two spins on the $A$- and $B$-sublattices point generally in the 
direction of the external magnetic field. However, they 
form angles $\alpha$ and $\pi-\alpha$ to the $x$-axis, as shown 
in the model state I of Fig.~\ref{model_states}b. The remaining spins 
on the $C$-sublattice point in a direction antiparallel to the 
applied external field. The second model state II of Fig.~\ref{model_states}c 
for the triangular lattice has two spins on the $A$- and $B$-sublattices 
that align completely with the external magnetic field and the remaining 
spins that align antiparallel to the external magnetic field. 
The final model state III has two spins on the $A$- and $B$-sublattices 
that form an angle $\alpha$ to the $x$-axis and another spin 
on the $C$-sublattice that forms a (initially negative) angle 
of $\beta$ to the $x$-axis, as is also shown in Fig.~\ref{model_states}d. 
Model state II is clearly a limiting case of both model states, 
I and III. (For example, we obtain model state II from model state 
III by setting $\alpha = \pi/2$ and $\beta = -\pi/2$.)

\begin{figure}
\epsfxsize=8cm
\centerline{\epsffile{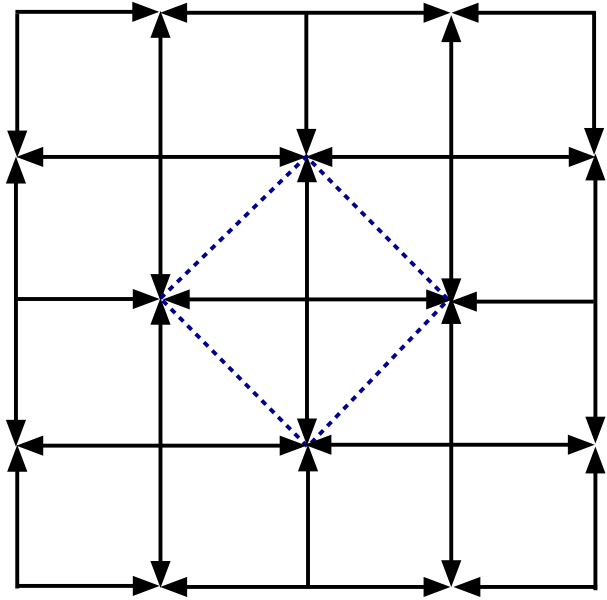}}
\caption{The bond directionality of the Heisenberg Hamiltonian after rotation of the local
coordinate axes
in the spin space. The directions of the bonds are indicated by the arrows placed on the square lattice. 
The two-site unit cell is also shown in dotted lines.}
\label{bond_directionality}
\end{figure}

In order to simplify the problem, we now rotate the local coordinate axes in
the spin space  
so that all spins appear notationally to point in the 
downwards $z$-direction. For a spin making an angle of $\theta$ 
to the $x$-axis, the rotation of the local axes is given by,
\begin{eqnarray}
s^x &\rightarrow&  - s^x {\rm sin}(\theta) + s^z {\rm cos}(\theta) \nonumber \\
s^y &\rightarrow&    s^y \nonumber \\
s^z &\rightarrow&  - s^x {\rm cos}(\theta) - s^z {\rm sin}(\theta) \;.\label{rotation}
\end{eqnarray}
The spins in the model state $|\Phi\rangle$ now all appear to point
downwards, i.e. $|\Phi\rangle=|\cdots \downarrow\downarrow\downarrow\downarrow
\cdots\rangle$. The corresponding creation  $\{C_I^{+}\}$ are then given by
$\{C_I^{+}\}
= { s}_{i}^+ \, , \, { s}_{i}^+{ s}_{j}^+ \, , \,
{ s}_{i}^+{ s}_{j}^+{ s}_{k}^+ \, ,\, \ldots \;$ ,
where the indices ${i},{j},{k},\ldots$ denote arbitrary lattice
sites.
Furthermore, the Hamiltonian for the
bipartite lattices in the rotated coordinate frame (i.e., with spins on the $A$
sublattice making an angle $\theta$ to the negative $x$-axis and spins 
on the $B$ sublattice making an angle $\theta$ to the positive 
$x$-axis as shown in Fig.~\ref{model_states}) is now given by
\begin{eqnarray}
H &=& \sum_{\langle i \rightarrow j \rangle} \biggl \{  
- \frac  14 (1+{\rm cos}(2\theta)) ( s_i^+ s_j^+ + s_i^- s_j^- )
\nonumber \\ & & ~~~~~~~
+\frac 14 (1-{\rm cos}(2\theta)) ( s_i^+ s_j^- + s_i^- s_j^+ ) 
\nonumber \\ & & ~~~~~~~
-  {\rm cos}(2\theta) s_i^z s_j^z + \frac 12 {\rm sin}(2\theta) ( s_i^z s_j^+ + s_i^z s_j^- )
\nonumber \\ & & ~~~~~~~
- \frac 12 {\rm sin}(2\theta) ( s_i^+ s_j^z + s_i^- s_j^z )
 \biggr \}
\nonumber \\ &+&
    \lambda {\rm sin}(\theta)   \sum_i  s_i^z 
    - \frac \lambda 2 {\rm cos}(\theta)  \sum_{i_A}  (s_{i_A}^+ + s_{i_A}^-) 
\nonumber \\ &+& 
  \frac \lambda 2 {\rm cos}(\theta)  \sum_{i_B}  (s_{i_B}^+ + s_{i_B}^-) \;.\label{rotH}  
\end{eqnarray}
We note that the sign in Eq. (\ref{rotH}) for those terms for $s^z s^+$ and 
$s^z s^-$ for a bond going from $i$ to $j$ has an opposite sign for those 
same terms for a bond going from $j$ to $i$. This is called  a ``bond directionality'' 
and is indicated in the above equation by the arrow in the symbol $\langle i 
\rightarrow j \rangle$. An illustrative example of bond directionality in 
the Hamiltonian for the square-lattice case is shown in Fig. \ref{bond_directionality}.
We note also that $i_A$ runs over all $A$ sublattice 
sites, $i_B$ runs over all $B$-sublattice sites, and $i$ runs over all lattice sites.  
The translational symmetry of Eq. (\ref{rotH}) compared to the original problem 
has also been reduced. We must include two sites in the unit cell, as
is also shown in Fig. \ref{bond_directionality}. 

Similar calculations may be carried out for the triangular lattice.
We have three new Hamiltonians after rotation of the local spin axes 
of the spins for all three model states I, II, and III in Fig.~\ref{model_states}(b-d) 
for the triangular lattice case such that all spins again appear to point downwards. 
The Hamiltonian for model state I, Fig.~\ref{model_states}(b), for the triangular 
lattice is:
\begin{eqnarray}
H &=& 
\sum_{\langle i_A \rightarrow i_B \rangle} \biggl \{  
- \frac  14 (1+{\rm cos}(2\alpha)) ( s_{i_A}^+ s_{i_B}^+ + s_{i_A}^- s_{i_B}^- ) 
\nonumber \\ & & ~~~~~~~
+ \frac 14 (1-{\rm cos}(2\alpha)) ( s_{i_A}^+ s_{i_B}^- + s_{i_A}^- s_{i_B}^+ ) 
\nonumber \\ & & ~~~~~~~
-  {\rm cos}(2\alpha) s_{i_A}^z s_{i_B}^z 
+ \frac 12 {\rm sin}(2\alpha) ( s_{i_A}^z s_{i_B}^+ + s_{i_A}^z s_{i_B}^- )
 \nonumber \\ & & ~~~~~~~
- \frac 12 {\rm sin}(2\alpha) ( s_{i_A}^+ s_{i_B}^z + s_{i_A}^- s_{i_B}^z )
 \biggr \} \nonumber \\
    &+&    \sum_{\langle i_{B,C} \rightarrow i_{C,A} \rangle} \biggl \{  
- \frac  14 (1+{\rm sin}(\alpha)) ( s_{i_{B,C}}^+ s_{i_{C,A}}^+ + s_{i_{B,C}}^- s_{i_{C,A}}^- ) 
\nonumber \\ & & ~~~~~~~
+ \frac 14 (1-{\rm sin}(\alpha)) ( s_{i_{B,C}}^+ s_{i_{C,A}}^- + s_{i_{B,C}}^- s_{i_{C,A}}^+ ) 
\nonumber \\ & & ~~~~~~~
-  {\rm sin}(\alpha) s_{i_{B,C}}^z s_{i_{C,A}}^z + \frac 12 {\rm cos}(\alpha) ( s_{i_{B,C}}^z s_{i_{C,A}}^+ + s_{i_{B,C}}^z s_{i_{C,A}}^- )
\nonumber \\ & & ~~~~~~~
- \frac 12 {\rm cos}(\alpha) ( s_{i_{B,C}}^+ s_{i_{C,A}}^z + s_{i_{B,C}}^- s_{i_{C,A}}^z )
 \biggr \} \nonumber \\              
  & &
  \nonumber \\ &-& 
  \lambda \sum_{i_C}  s_{i_C}^z 
    + \lambda {\rm sin}(\alpha) (  \sum_{i_A} s_{i_A}^z +  \sum_{i_B}    s_{i_B}^z ) 
    \nonumber \\ &-& 
  \frac {\lambda}2 {\rm cos}(\alpha)  \sum_{i_A}  (s_{i_A}^+ + s_{i_A}^-)   
     + \frac {\lambda}2 {\rm cos}(\alpha)  \sum_{i_B}  (s_{i_B}^+ + s_{i_B}^-)\;,\label{rotH2}     
\end{eqnarray}
where the sum $\langle i_A \rightarrow i_B \rangle$ goes from sublattice $A$ to 
sublattice $B$ (and with directionality). Note that  $\langle i_{B,C} \rightarrow i_{C,A} 
\rangle$ indicates a sum that goes from sublattice $B$ to sublattice 
$C$ and sublattice $C$ to sublattice $A$, respectively (and with directionality). A similar
treatment may be carried out for the model state III, Fig.~\ref{model_states}(d). Hence, if
those spins on on the $A$ and $B$ sublattices make an angle $\alpha$ to 
the $x$-axis and those spins on the $C$ sublattice make an angle $\beta$ to the 
$x$-axis and employing the rotation of the local spin axes of Eq. (\ref{rotation}), 
we find that,
\begin{eqnarray}
H &=& 
 \sum_{\langle i_C \rightarrow i_{A,B} \rangle} \biggl \{  
 \frac  14 (-1+{\rm cos}(\alpha-\beta)) ( s_{{i_C}}^+ s_{{i_{A,B}}}^+ + s_{{i_C}}^- s_{{i_{A,B}}}^- ) \nonumber \\
& & ~~~~~~~
+ \frac 14 (1+{\rm cos}(\alpha-\beta)) ( s_{{i_C}}^+ s_{{i_{A,B}}}^- + s_{{i_C}}^- s_{{i_{A,B}}}^+ )
\nonumber \\ & & ~~~~~~~
    +  {\rm cos}(\alpha-\beta) s_{{i_C}}^z s_{{i_{A,B}}}^z 
\nonumber \\ & & ~~~~~~~
    + \frac 12 {\rm sin}(\alpha-\beta) ( s_{{i_C}}^+ s_{{i_{A,B}}}^z + s_{{i_C}}^- s_{{i_{A,B}}}^z )
\nonumber \\ & & ~~~~~~~
   - \frac 12 {\rm sin}(\alpha-\beta) ( s_{{i_C}}^z s_{{i_{A,B}}}^+ + s_{{i_C}}^z s_{{i_{A,B}}}^- ) \biggr \} 
\nonumber \\ &+& 
   \sum_{\langle i_A , i_B \rangle} \biggl \{  
               \frac 12 ( s_{i_A}^+ s_{i_B}^- + s_{i_A}^- s_{i_B}^+ ) + s_{i_A}^z s_{i_B}^z  \biggr \} 
      \nonumber \\
     &+& 
    \lambda {\rm sin}(\alpha) (   \sum_{i_A} s_{i_A}^z +  \sum_{i_B}    s_{i_B}^z )
    + \lambda {\rm sin}(\beta) \sum_{i_C}  s_{i_C}^z 
      \nonumber \\
     &+& 
    \frac {\lambda}2 {\rm cos}(\alpha) \{ \sum_{i_A}  (s_{i_A}^+  + s_{i_A}^-)  
    + \sum_{i_B}  (s_{i_B}^+ + s_{i_B}^-) \} 
\nonumber \\ &+& 
    \frac {\lambda}2 {\rm cos}(\beta)  \sum_{i_C}  (s_{i_C}^+ + s_{i_C}^-) \; ,  
    \label{rotH3} 
\end{eqnarray}
where the sum $\langle i_C \rightarrow i_{A,B} \rangle$ goes from sublattice $C$ to 
sublattices $A$ and $B$ (with directionality) and $\langle i_A , i_B \rangle$ goes over 
each bond connecting the $A$ and  $B$ sublattices, but counting each one once only 
(and without directionality). We note that we have three sites in the unit cell for 
all of the models states used for the triangular lattice antiferromagnet. 

Note that in addition to the model states presented above, 
spin liquids such as valence-bond crystal states may be treated via the
CCM is by using a dimerised or plaquette (etc.) as relevant model state. 
A corresponding matrix algebra \cite{ccm5} is then used with respect to this state. 
However, a simpler approach is now also available that relies on finding 
special solutions of the CCM equations for the N\'eel-type model 
states used here \cite{ccm35}. These allow us to treat via existing 
high-order formalism and computer code, for example, spontaneous 
symmetry breaking in the spin-half one-dimensional $J_1$--$J_2$ 
(Majumdar-Ghosh) model \cite{ccm35}. The CCM is thus not restricted 
purely to semi-classical systems.

We consider the angles as free parameters in the CCM calculation. 
They are determined by direct minimisation 
of the CCM ground-state energy. This was achieved computationally at a given level of
LSUB$m$ approximation, and a minimum ground state energy with respect to these 
canting angles was also found computationally for a given fixed value of $\lambda$. 
There was only one angle for the square-lattice antiferromagnet (and for model state 
I for the triangular lattice) and there were two such angles for model state III for the 
triangular lattice.  The next value of $\lambda$ was then determined incrementally 
and the minimisation process of the energy with respect to the canting angles repeated. 
The fact that we had to minimise the ground-state energy with respect to such 
angles at each value of $\lambda$ made the CCM calculations much more costly 
in terms of computing time required than the equivalent situations at zero external 
magnetic field, which requires no such minimisation. Furthermore, we see
that the Hamiltonians of Eqs. (\ref{rotH}-\ref{rotH3}) do not conserve the quantity
$s_T^z \equiv \sum_i s_i^z = 0$, which is preserved for the square-lattice 
antiferromagnet at $\lambda=0$. For these reasons, CCM calculations in 
the presence of external magnetic fields are more challenging than their
zero-field counterparts. 

A final point is that 
the inclusion of the CCM SUB1 terms of form $S_1 \equiv {\cal S}_{i_1} s_i^+$ in 
the ground ket and bra states is also equivalent to a rotation of the local spin axes 
\cite{ccm1}. For example, for the spin-half system, we note that $(s_i^+)^2 | 
\Phi\rangle$=0 and so we can prove that $e^{S_1}|\Phi \rangle={\Pi}_i 
(1+{\cal S}_{i_1} s_i^+)|\Phi\rangle$. This produces a mixture of ``up" and ``down" 
spins at each site, which may be thought of (as may be seen from Eq.~(\ref{rotation}) 
above, for example) as the same as a rotation of local spin axes. Hence, 
we conclude that SUB1 is equivalent to a rotation of the axes. Previous 
calculations for Heisenberg antiferromagnets in external magnetic fields \cite{ccm1}
made the explicit assumption that the correlation coefficients of the SUB1 
terms may be set to zero,
%and then use the SUB1 equations as a consistency 
%equation in order to find the canting angles used in the model state. 
and we make the same explicit assumption here. We minimise the 
ground-state energy explicitly with respect to the angles in our model
state. Note that we go to much higher orders of LSUB$m$ approximation
than those calculations presented in \cite{ccm1}.

%Again, the manner in which the CCM ket- and bra-state equations may be 
%obtained computationally has been discussed extensively elsewhere 
%\cite{ccm12,ccm15,ccm20,ccm26}, and so this is not considered in detail here. 
%However, we note that the determination of the ket-state equations 
%reduces essentially to a pattern-matching exercise 
%of spin-lowering operators in $C_i^-$ to the similarity transformed 
%version of the Hamiltonian. Furthermore, the bra-state equations are 
%solved easily once the solution to the ket-state equations has been 
%obtained. This is straightforward to carry out and this process is 
%discussed in more detail in Refs.  \cite{ccm12,ccm15,ccm20,ccm26}. 

To investigate the magnetisation process in antiferromagnets we have to
consider  
the total lattice magnetisation $M$ along the direction of the magnetic
field.
This quantity (in the initial coordinate frame prior to rotation of the local spin axes) 
is defined by $M =  
\frac{1}{Ns} \langle  \sum_i  s_i^z \rangle=\frac{1}{Ns} \langle \tilde \Psi | 
\sum_i  s_i^z | \Psi \rangle$ ($s$ is the spin quantum number which is
$s=1/2$ throughout this paper).
In the rotated coordinate frame (and in which  
all of the spins point appear ``mathematically" to downwards), the 
lattice magnetisation for the bipartite lattices is now given by
\begin{eqnarray}
M &=& -\frac{{\rm sin}(\theta)}{Ns} \sum_i \langle \tilde \Psi |  s_i^z | \Psi \rangle
   - \frac{{\rm cos}(\theta) }{2Ns}  \sum_{i_A} \langle \tilde \Psi |   s_{i_A}^+ + s_{i_A}^- | \Psi \rangle  
\nonumber \\ & & ~~~~~~~
   + \frac{{\rm cos}(\theta)}{2Ns}  \sum_{i_B}  \langle \tilde \Psi | s_{i_B}^+ + s_{i_B}^- | \Psi \rangle \; ,
 \label{rotM}  
\end{eqnarray}
where, again, $i_A$ runs over all $A$ sublattice sites, $i_B$ runs over 
all $B$-sublattice sites, and $i$ runs over all lattice sites. We are 
able to determine readily the lattice magnetisation once the ket- and 
bra-state equations have been solved for a given value of $\lambda$. 
Furthermore, similar expressions to Eq.~(\ref{rotM}) may be obtained 
for the lattice magnetisation for the triangular lattice for model 
states I, II, III, Fig.~\ref{model_states}(b)-(d). We note that the 
magnetisation found on the three sublattices may become non-equivalent in 
a magnetic field for the triangular-lattice case. Indeed, for the 
triangular lattice, the expression for the lattice magnetisation aligned in
the direction of the applied magnetic field 
on the individual sublattices (denoted, $M_A$, $M_B$, and $M_C$) in 
terms of the global axes prior to rotation of the local spin axes 
is given by
\begin{equation}
M_{A,B,C} =
   \frac 1{N_{A,B,C}\;s}   \sum_{i_{A,B,C}}   
      \langle \tilde \Psi | s_{i_{A,B,C}}^z | \Psi \rangle ~~ ,
 \label{rotMabc}  
\end{equation}
where the index $i_a$ runs over all $N_A$ sites on sublattice $A$,  
the index $i_B$ runs over all $N_B$ sites on sublattice $B$, and the 
index $i_C$ runs over all $N_C$ sites on sublattice $C$. Clearly, 
we see that $N=N_A+N_B+N_C$ and that $M=(M_A+M_B+M_C)/3$.

\begin{figure}
\epsfxsize=11cm
\centerline{\epsffile{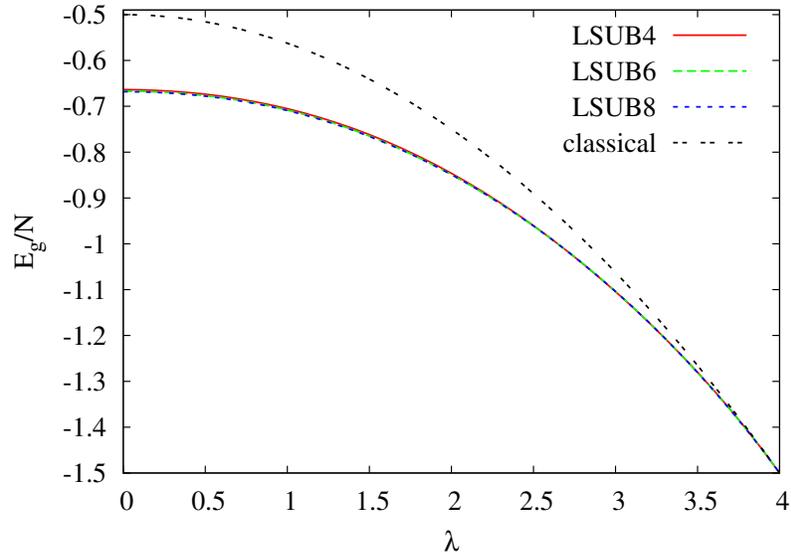}}
\caption{Results for the ground-state energy per site $E_g/N$ of the spin-half 
square-lattice Heisenberg antiferromagnet in dependence on an external 
magnetic field of strength $\lambda$. Note that the curves for LSUB4, LSUB6, LSUB8 almost coincide.}
\label{square_energy}
\end{figure}

\begin{figure}
\epsfxsize=11cm
\centerline{\epsffile{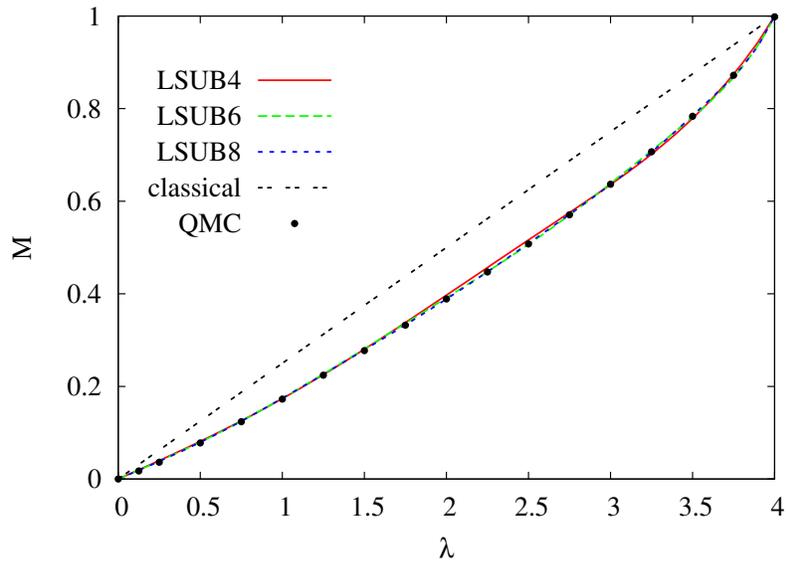}}
\caption{Results for the total lattice magnetisation $M$ 
of the spin-half square-lattice Heisenberg
antiferromagnet in 
the presence of an external magnetic field of strength $\lambda$ compared to results of 
QMC \cite{squareTriangleED}. Note that the curves for LSUB4, LSUB6, LSUB8 almost coincide.}
\label{square_magnetisation}
\end{figure}

\begin{figure}
\epsfxsize=11cm
\centerline{\epsffile{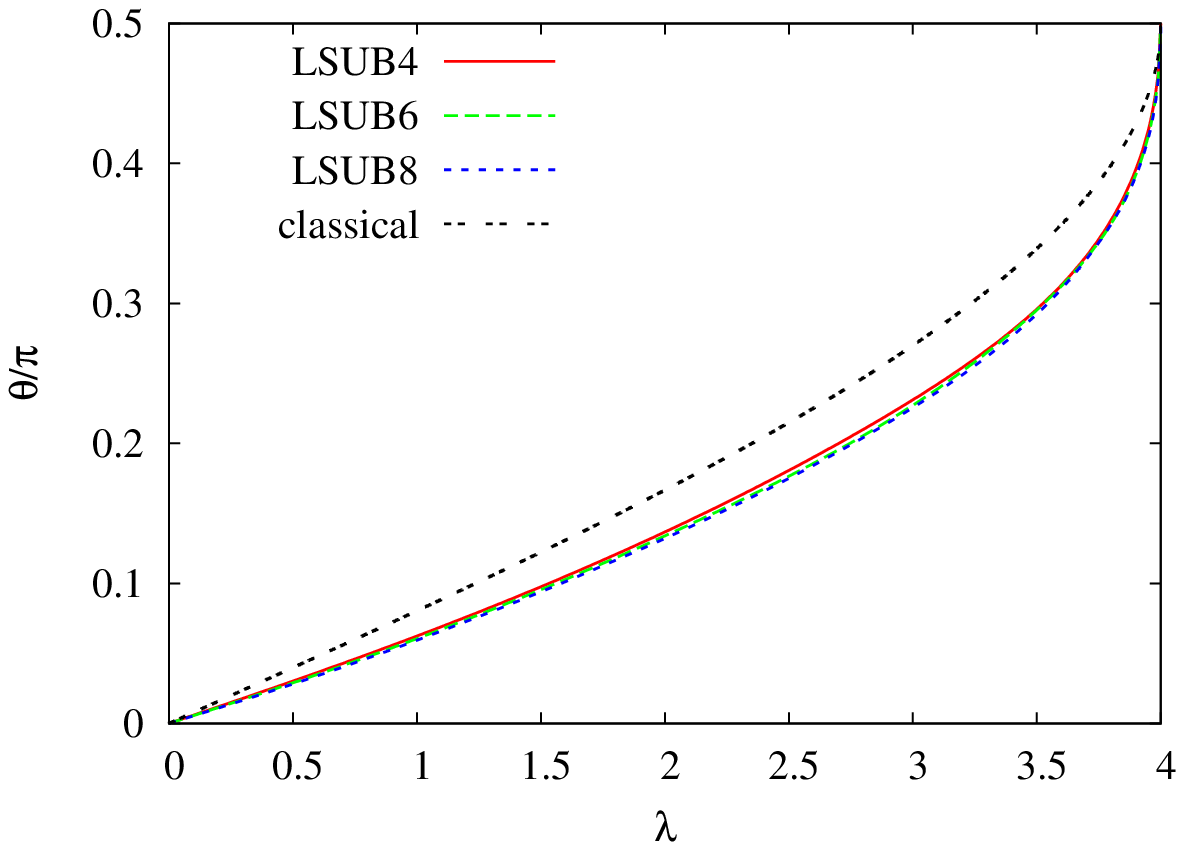}}
\caption{Results for the canting angle $\theta/\pi$ obtained for the model state for 
the spin-half square-lattice  Heisenberg
antiferromagnet (see Fig.~\ref{model_states}a) in the presence of an external magnetic 
field of strength $\lambda$. Note that the curves for LSUB4, LSUB6, LSUB8 almost coincide.}
\label{square_angle}
\end{figure}

\section{Results}
Now we present and discuss the results for the two models under
consideration calculated by the CCM as illustrated above.
We start with the spin-half square-lattice Heisenberg
antiferromagnet. 
The ground-state energy in dependence of this model
is shown in Fig.~\ref{square_energy}.  The CCM results 
converge rapidly with increasing LSUB$m$ level of approximation. 
As seen in previous CCM calculations \cite{ccm15}, the ground-state 
energy in the limit of vanishing external field ($\lambda=0$) is approximated 
well. The interested reader is referred to Refs. \cite{ccm15} for 
a more detailed discussion of these results. We also find that the exact 
result for the saturation field $M=M_s$ at $\lambda_s=4$ is also reproduced. 
At this point the spins all lie in the direction of the external field.

The results for the lattice magnetisation are shown in Fig.~\ref{square_magnetisation}. 
There is a considerable difference between the results for the spin-half quantum model 
and the classical straight-line behaviour (i.e., $M_{\rm{Classical}}=\frac{1}{4}\lambda$). 
Clearly, this difference is because of quantum effects. It is also obvious 
from Fig.~\ref{square_magnetisation} that the magnetisation of the quantum model
is below that of the classical magnetisation in the region $0 < \lambda < \lambda_s$.  
Again we note that the LSUB$m$ results appear to converge with increasing $m$ 
for all values of $\lambda$. 
For example, the difference between 
the  LSUB6 and LSUB8 results for the lattice magnetisation 
is less than $2\cdot10^{-3}$ for all values of $\lambda$, and it is impossible to
be detected by eye in Fig.~\ref{square_magnetisation}.
From  Fig.~\ref{square_magnetisation} it is also evident that the CCM results 
for the lattice magnetisation are in excellent agreement with the results of QMC \cite{squareTriangleED},
which can be considered as the most accurate results available.

In addition to the energy and the magnetisation we can also present 
results for the canting angle $\theta$ (cf. Fig.~\ref{model_states}) 
of the quantum model, see Fig.~\ref{square_angle}. Again, 
there is a noticeable difference between the values for the classical 
and the quantum angle. This difference first increases with $\lambda$ up to
about $\lambda \approx 3.5$. Beyond $\lambda \approx 3.5$ the quantum angle
very rapidly approaches the saturation value $\theta_s = \pi/2$.

In the next step the CCM results for the ground state energy and the lattice 
magnetisation in dependence on magnetic field can be used 
to 
calculate the uniform magnetic susceptibility, given by
\begin{equation} \label{susc}
\chi \equiv \frac{1}{2}\frac{dM}{d\lambda} = 
- \frac{1}{N}\frac{d^2 E_g}{d\lambda^2} \; .
\end{equation}
Note that factor of $\frac{1}{2}$ in $\frac{1}{2}\frac{dM}{d\lambda}$ is due to definition of
$M$ in the interval $[0,1]$.
Note further that we consider here $\chi$ as susceptibility per site
\cite{defin_chi}. 
For the concrete calculation of $\chi$ we have used the second derivative of the
energy. To check the accuracy for low fields  
we have also determined $\chi$  numerically via direct determination from $M$ by using 
$\frac{dM}{d\lambda}$. We found that $\frac{1}{2}\frac{dM}{d\lambda}$ and
$\frac{1}{N}\frac{d^2 E_g}{d\lambda^2}$  agree to at least six decimal places of precision. 

\begin{figure}
\epsfxsize=11cm
\centerline{\epsffile{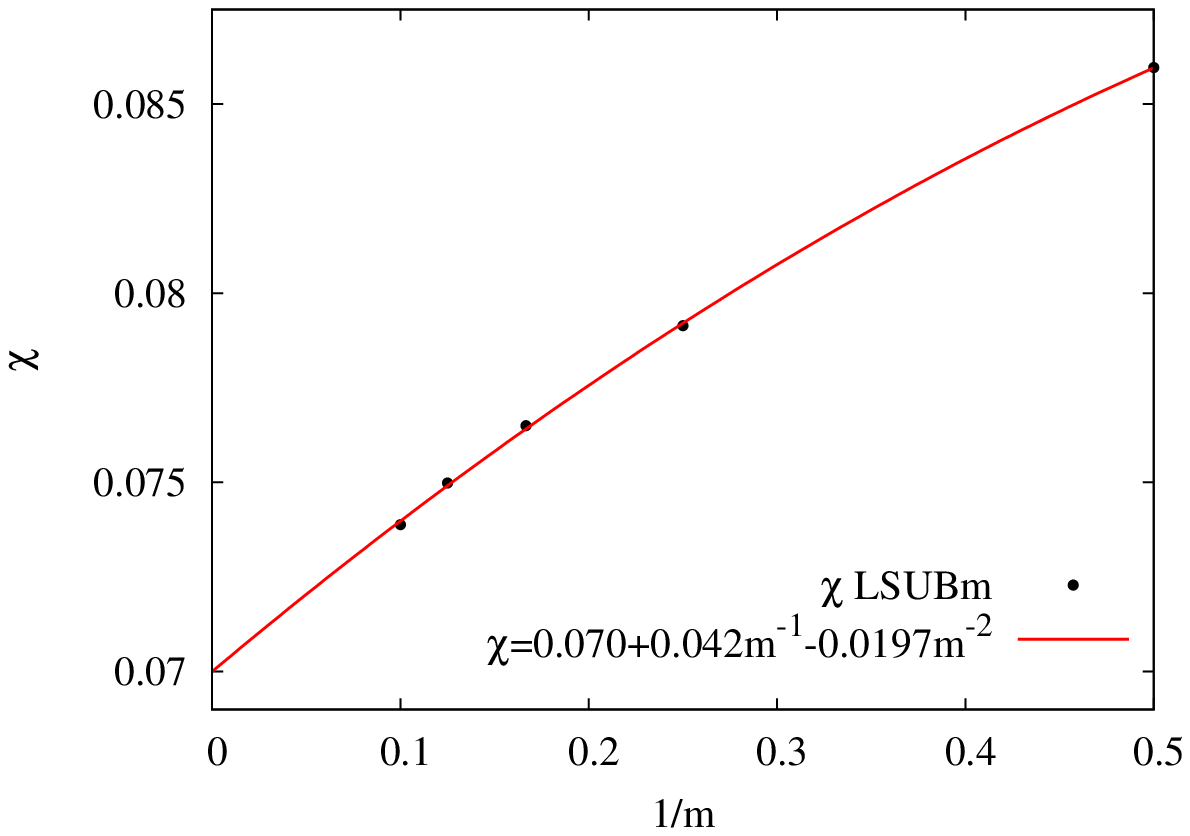}}
\caption{LSUB$m$ results for the 
zero-field uniform susceptibility $\chi({\lambda \rightarrow 0})$
for the spin-half square-lattice  Heisenberg
antiferromagnet (see Fig.~\ref{model_states}a) with $m=\{2,4,6,8,10\}$ and
the polynomial fit according to $\chi(m)=c_0+c_1/m+c_2/m^2$.}
\label{extrap_square_chi}
\end{figure}

The zero-field uniform susceptibility $\chi({\lambda \rightarrow 0})$, the ground state energy, the sublattice 
magnetisation, the spin stiffness, and the spin-wave velocity constitute the 
fundamental parameter set that determines the low-energy physics of
magnetic systems. The results for the ground state energy, the 
sublattice magnetisation, the spin stiffness for the square-lattice Heisenberg
antiferromagnet at $\lambda=0$ have been calculated by the CCM previously. 
The interested reader is referred to Refs. \cite{ccm15}
for more details. However, CCM results for the susceptibility $\chi$ 
were not determined by these earlier calculations. 
Here we find that 
$\chi$=$0.08596$, $0.07915$, $0.07650$, $0.07498$, and $0.07388$ 
for the LSUB2, LSUB4, LSUB6, LSUB8, and LSUB10 approximations, respectively. 
Since the LSUB$m$ approximation becomes exact for $m \to \infty$,
it is useful to extrapolate the ``raw'' LSUB$m$ data to $m \to \infty$.
Meanwhile there is much empirical experience
how to extrapolate CCM LSUB$m$ data for physical quantities such as the spin stiffness
\cite{ccm27,ccm32} and ``generalised'' susceptibilities
\cite{ccm32} which are also related 
to a second derivative of the
ground  energy $E_g$.
Hence, we use
the same extrapolation rule for 
the zero-field uniform susceptibility
that has
previously been found to give good results for the spin stiffness 
and also for ``generalised'' susceptibilities \cite{ccm27,ccm32} 
given by 
$\chi(m)=c_0+c_1/m+c_2/m^2$. % (polynomial). 
We see from Fig.  \ref{extrap_square_chi} that this rule provides a 
good method of extrapolation of our data. The corresponding extrapolation then yields
values for the susceptibility of $\chi=0.0700(6)$.  (The number in 
brackets indicate the standard deviation.) 
This result is in reasonable  
agreement with data obtained by other methods, e.g  QMC  ($\chi =
0.0669(7)$) \cite{runge}, series expansion ($\chi = 0.0659(10)$) \cite{zheng},
linear spin-wave theory ($\chi = 0.05611$) \cite{lswt},  
second-order spin-wave theory ($\chi = 0.06426$) \cite{square1},  and 
third-order spin-wave theory ($\chi = 0.06291$) \cite{hamer}. 
%We believe that even closer agreement would occur with high orders
%of LSUB$m$ approximation.
 
\begin{figure}
\epsfxsize=11cm
\centerline{\epsffile{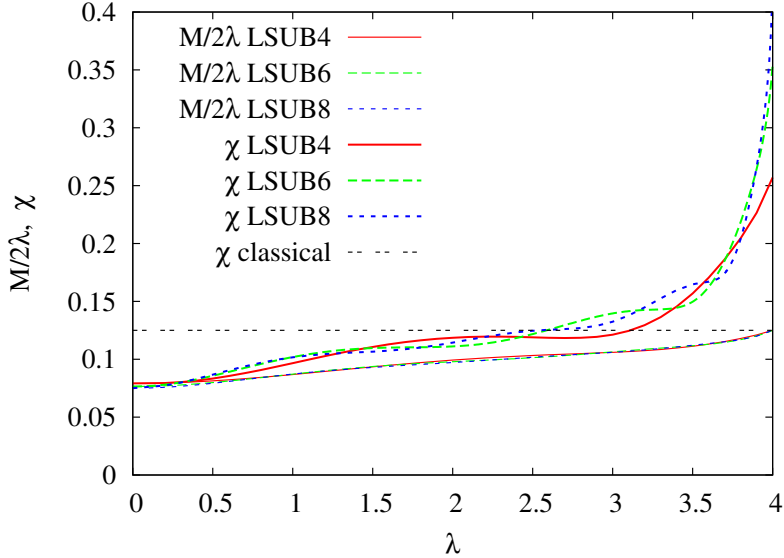}}
\caption{Susceptibility $\chi$, see Eq.
(\ref{susc}), and  the quotient $M/2\lambda$ in dependence on the magnetic
field $\lambda$ the for the spin-half square-lattice  Heisenberg
antiferromagnet. Note that the $M/2\lambda$ curves for LSUB4, LSUB6, LSUB8 almost
coincide.}
\label{chi_vs_H_sq}
\end{figure}

The field dependence of $\chi$ is also of experimental 
interest, see e.g. \cite{schroeder,zheludov,schroeder1,luban}. 
We present LSUB4, LSUB6, and LSUB8 data for the field dependence
of $\chi$  in Fig.~\ref{chi_vs_H_sq}. We note that the magnetisation divided by the 
applied external field is often considered in experimental studies. 
Hence, results for $M/2\lambda$ are given also in Fig.~\ref{chi_vs_H_sq}. 
For the sake of comparison, the classical value $\chi_{\rm clas}=1/8$ 
is also shown in this figure and we remark that  this value is clearly 
independent of  $\lambda$. From Fig.~\ref{chi_vs_H_sq} it is obvious that
$\chi$ and $M/2\lambda$ agree well with each other up to about
$\lambda=0.4=\lambda_s/10$. The difference between results of 
the LSUB8 approximation and the classical result is about 4\% at 
$\lambda=0.4$). However, these two sets of results begin to deviate 
significantly for larger $\lambda$. Hence,  the quantity $M/2\lambda$
is a good approximation for $\chi$ for magnetic fields used in real
experiments for systems with large saturation fields $\lambda_s$, 
and {\it not} for systems with low $\lambda_s$.
We observe  that $\chi$  increases with $\lambda$ as we move away 
from the zero-field point, $\lambda=0$. 
Similar increases in $\chi$ with the external field have been observed
experimentally, e.g.,  for the quasi-two-dimensional antiferromagnet
Ba$_2$CuGe$_2$O$_7$ \cite{zheludov}. 
Moreover, these results are in agreement with recent results obtained by 
exact diagonalisations, QMC simulations,  and spin-wave theory 
\cite{luscherlauchli,zhito09}. 
As seen for these other methods, 
the susceptibility is near the constant classical value for magnetic fields 
$1.5 \lesssim \lambda \lesssim 3.5$, although it starts rapidly to increase
approaching the saturation field. Finally, weak oscillations seen for 
$1.5 \lesssim \lambda \lesssim 3.5$, although these are believed to be 
artefacts of CCM LSUB$m$ approximation. We note that the number 
of oscillations increases are we increase the LSUB$m$ approximation
level, although their amplitude decreases markedly. In the limit, $m
\rightarrow \infty$, it is expected that these oscillations will disappear 
entirely.

We conclude from all of these results that the CCM provides precise results for 
the behaviour of the spin-half square-lattice quantum antiferromagnet in 
an external magnetic field. However, we see also from these results that 
the classical picture is essentially correct. Quantum mechanical effects 
modify, but do not change, the essential physics that occur in this 
unfrustrated quantum spin system. 

We now consider the spin-half antiferromagnet on the  triangular lattice. 
However, the situation is more complicated here because we have three 
sublattices in this case. As discussed above, we employ therefore the 
model states I, II, III shown in Fig.~\ref{model_states}(b-d).  
The computational effort of the CCM calculations presented here 
for the model state III to very high orders is very great because we also 
need to find the minimum of the energy with respect to two canting angles,
namely $\alpha$ and $\beta$.
The CCM calculation for the model state III in LSUB8
approximation was performed on a Beowulf cluster using 110
cores (Intel XEON 3GHz CPU). On this computer the running time for one data point
was approximately 2 days. 
The CCM has been shown to be fully competitive with
the results of other methods at the levels of approximation currently
available to use using parallel computer methods (currently: a maximum 
of 1000 CPUs in parallel). The interested reader is referred, e.g., to Refs. 
\cite{ccm12,ccm15,ccm20,ccm26}  for detailed comparisons of CCM results 
to the best of other methods.
   
\begin{figure}
\epsfxsize=11cm
\centerline{\epsffile{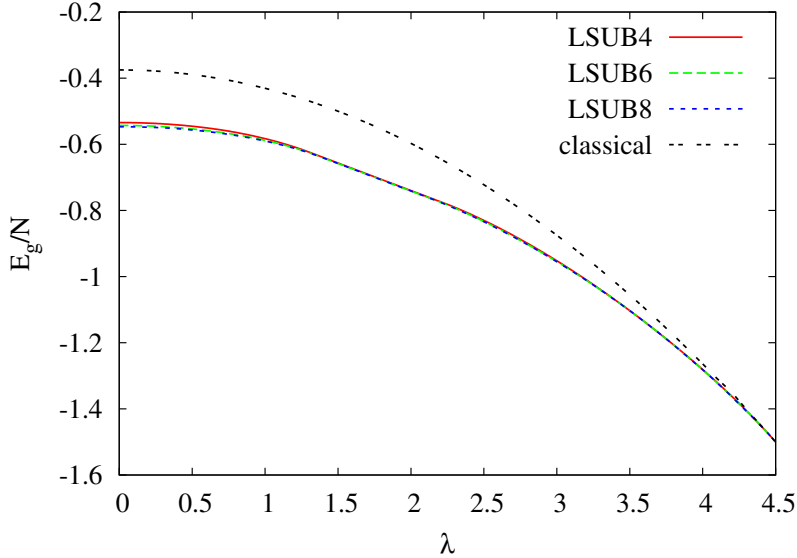}}
\caption{
Results for the ground-state energy per site $E_g/N$ 
of the spin-half triangular-lattice Heisenberg
antiferromagnet in the presence of an external magnetic field of strength $\lambda$.
Note that the results for LSUB4, LSUB6, LSUB8 are clearly converging 
rapidly for all values of $\lambda$.}
\label{triangle_energies}
\end{figure}

The results for the ground-state energy are shown in Fig.~\ref{triangle_energies}.
We note that the results for the model state with lowest energy 
are shown only as a function of $\lambda$ in Fig.~\ref{triangle_energies}.
Thus, results of model state I only are presented for small values of
the applied magnetic field strength $\lambda$ and results of model state 
III only are presented for higher values of $\lambda$ near to 
$\lambda_s$. The results of both model states coincide in the intermediate
regime. Again, these LSUB$m$ series of results are found to converge rapidly with 
increasingly levels of LSUB$m$ approximation over all values of the 
external field parameter $\lambda$. As may also be observed in 
Fig.~\ref{triangle_energies}, there is also a large reduction in 
the ground-state energy of the CCM results compared to the classical 
results for the energy (except in the trivial limit $\lambda \rightarrow 
\lambda_s=4.5$).

%\textcolor{red}{We remark also that results for model state III were found to 
%have poor convergence in the region $3.9 \le \lambda \le 4.3$. Results for LSUB8 
%in this region are therefore omitted from all figures for the triangular lattice model. We 
%believe that this problem with convergence might be rectified by a using smaller 
%step size for $\lambda$ and smaller initial changes in the values of $\alpha$ and 
%$\beta$ used in finding the minimum ground-state energy. However, this lies 
%beyond our current computational capabilities at the moment. Clearly though, 
%the results for the LSUB$m$ series are converging rapidly for all other values
%of $\lambda$ and we expect similar agreement for LSUB8 with results of lower
%orders of approximation in the region $3.9 \le \lambda \le 4.3$.}

\begin{figure}
\epsfxsize=11cm
\centerline{\epsffile{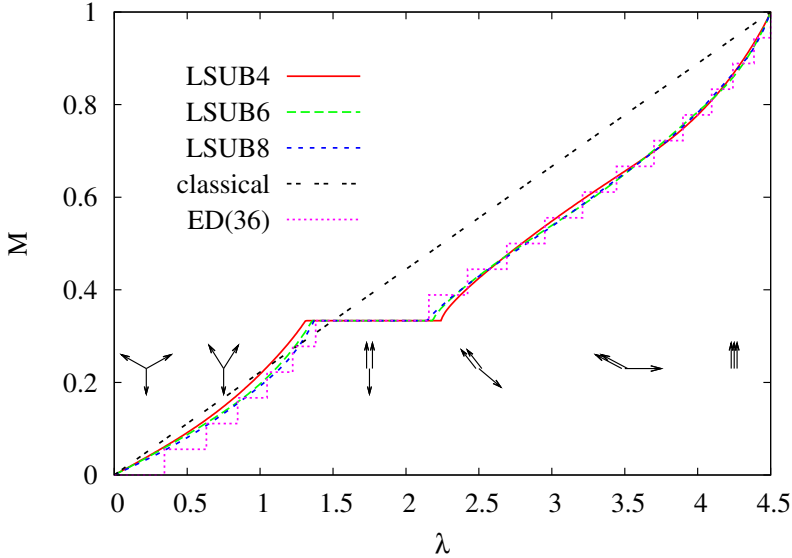}}
\caption{Results for the total lattice magnetisation $M$ 
of the spin-half triangular-lattice Heisenberg
antiferromagnet in the presence of an external magnetic field of strength $\lambda$. 
CCM results are compared to those results of exact diagonalisations 
\cite{squareTriangleED}. The arrows illustrate the actual spin directions.  We use model state I for 
$\lambda \le \lambda_1$ and we use model state III for $\lambda \ge \lambda_2$ (see Fig.~\ref{model_states}). Both model states give identical results within the plateau $\lambda_1 \le \lambda \le \lambda_2$ .}
\label{triangle_magnetisation}
\end{figure}

The results for the total lattice magnetisation are shown in 
Fig.~\ref{triangle_magnetisation}. 
The LSUB$m$ results are again seen 
to converge rapidly for increasing $m$. However, there is a 
radical departure from the classical straight-line behaviour (i.e.
$M_{\rm{Classical}}=\frac{2}{9}\lambda$) in this case. Thus, we find that the 
quantum model deviates from the linear relationship between $M$ 
and $\lambda$. The most prominent feature of our CCM results 
is the plateau in the $M$ versus $\lambda$ curve at $M/M_s=\frac 13$. 
Note that the plateau corresponds to the ``straight" part of the curve in 
the $E_g(\lambda)$ curve shown in Fig.~\ref{triangle_energies}.
Note further that this plateau is well-known and has been found by 
other approximate methods
\cite{Hon1999,HSR04,nishi,chub,ono,squareTriangleED}. 
The ground state of the quantum system over the finite, non-zero range 
of $\lambda$ for the plateau region has ordering of the form shown 
in model state
II of Fig.~\ref{model_states}(c). Importantly, this is an example of when 
quantum fluctuations favour collinear ordering (so called `order from
disorder' phenomenon, see e.g. Refs.~\cite{villain,shender,Kubo_JPSJ}).
This plateau state of 
model state II is observed only at a single point classically, namely, at
$\lambda=1.5$. The classical ground state is given by model state II 
in Fig.~\ref{model_states}(c) only at this point, see also
Ref.~\cite{kawamura,zhito,cabra}.  
Indeed, states I, II and III are equivalent classically at the point 
$\lambda=1.5$. 
The values for the starting ($\lambda_1$) and the end point ($\lambda_2 $) 
of the plateau state calculated within different LSUB$m$ approximations are 
shown in Table \ref{tab1}. 
The most accurate values are provided by the LSUB8 approximation, namely,
that $\lambda_1 \approx 1.37$ and $\lambda_2 \approx 2.15$. These results 
may therefore serve as the CCM estimate for  the plateau width. 
We note that the results for $\lambda_1$ and $\lambda_2$ for even and 
odd values of $m$ ought to converge to the same values in this limit.
% \cite{ccm28a}. 
Our estimate for the range of the plateau 
is in also reasonable agreement with those results of spin-wave theory 
\cite{chub} and exact diagonalisations \cite{squareTriangleED}, 
which both predict a similar width for the plateau with respect 
to the applied external magnetic field. 
However, we note that spin-wave theory was carried out only 
to order $1/S$ for the triangular lattice antiferromagnet in an 
external field. We believe that higher orders than $1/S$ for 
spin-wave theory would provide better correspondence to those 
results of ED and CCM results cited here regarding the range 
of the plateau.
The phenomenon of ``order from disorder'' in
which quantum fluctuations tend to favour colinear 
states has studied extensively elsewhere, e.g., Refs. 
\cite{villain,shender}. We note that the plateau state 
(uud) is colinear in the present case, and so our results 
are another example of this phenomenon. We have shown 
here that quantum fluctuations stabilise the (uud) state 
over other states that classically would have had lower 
energy in the plateau region.

\begin{figure}
\centerline{\epsfxsize=7.8cm\epsffile{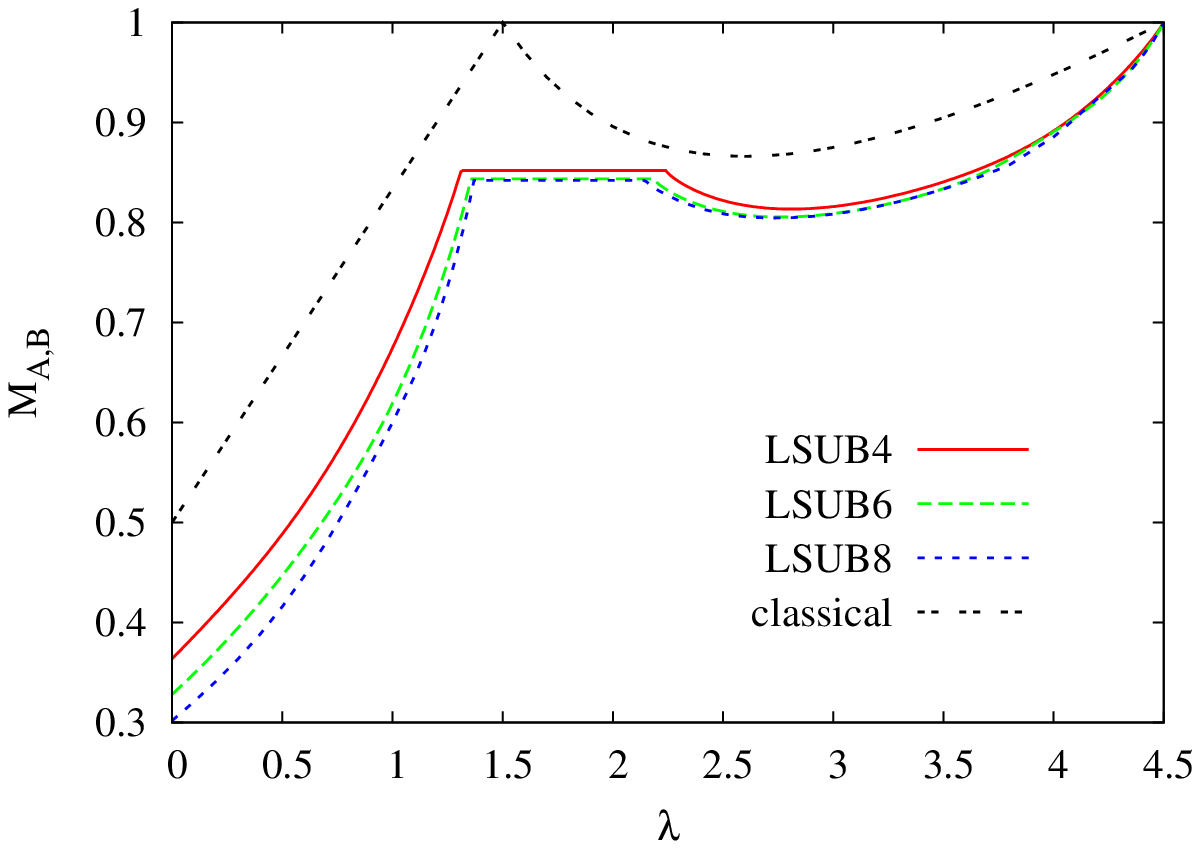} \hspace*{0.5cm}
            \epsfxsize=7.8cm\epsffile{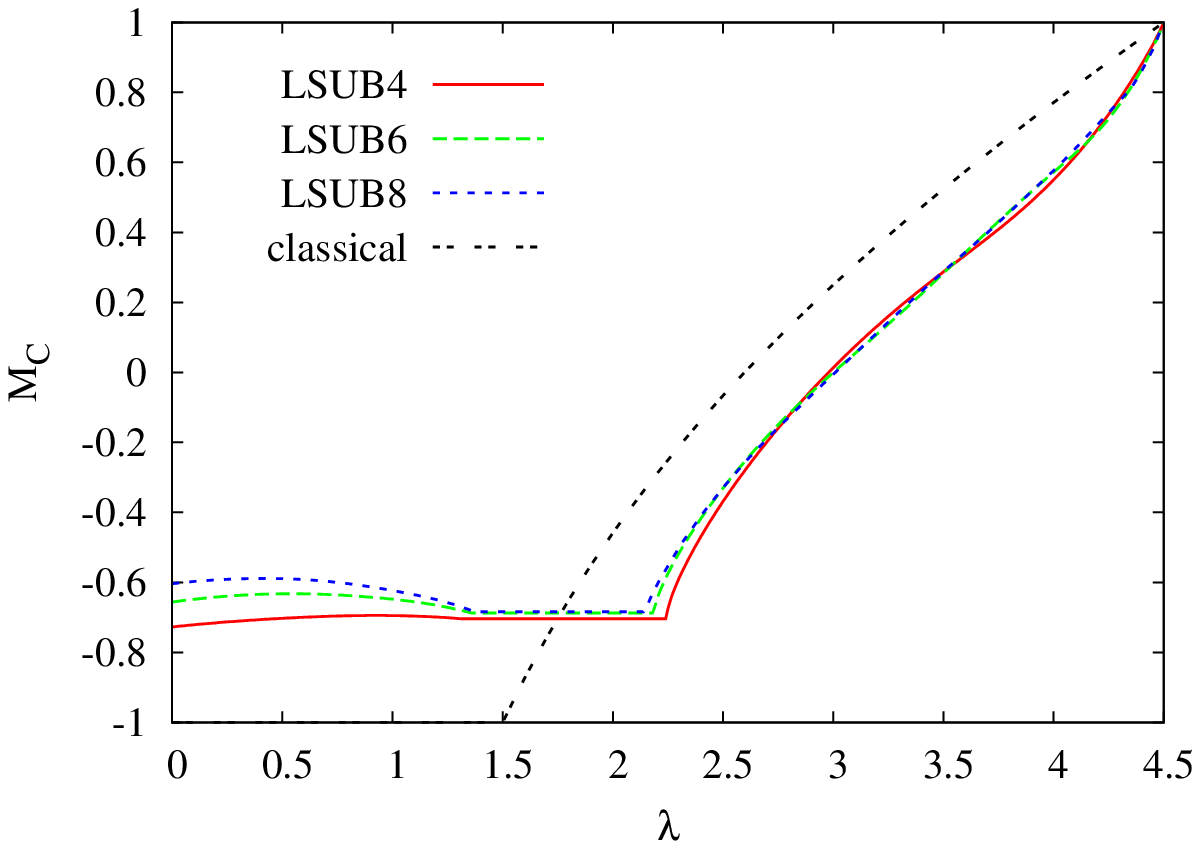}}
\caption{Results for the sublattice magnetisation $M_{\gamma}$ ($\gamma=\{A,B,C\}$) on 
individual sublattices $A$ and $B$ (left) and $C$ (right) of the spin-half 
triangular-lattice Heisenberg antiferromagnet in the presence of an external magnetic field of strength $\lambda$. 
(Note that $M_A=M_B$ for all $\lambda$.) }
\label{triangle_magnetisation_a_b_c}
\end{figure}

We are able also to calculate the (sub)lattice magnetisation (i.e., with respect to 
the $z$-direction in the original unrotated spin axes) for the individual sublattices, 
namely,  $M_A$, $M_B$ and $M_C$ given by Eq.~(\ref{rotMabc}), by using the CCM
and as a function of $\lambda$.   As far as we are aware, these quantities
have never before been presented for this model. The results for $M_A$, $M_B$ 
and $M_C$ are now presented in Fig.~\ref{triangle_magnetisation_a_b_c}. 
Once again, we see a radical shift in the quantum solution from the classical result. 
Interestingly, $M_C$ appears to decrease before approaching the plateau at
$\lambda=\lambda_1$, while $M_A=M_B$ increase monotonically 
with $\lambda$ up to $\lambda_1$.  On the other hand, $M_A, M_B$ decrease 
with magnetic field in the region $\lambda_2 < \lambda \lesssim 2.8$ 
above the plateau, while $M_C$  increases monotonically with $\lambda$ up to $\lambda_s$.   

\begin{table}[ph]
\caption{CCM results for the width of the magnetisation plateau for the spin-half Heisenberg antiferromagnet on the triangular lattice.}
\begin{center}
\begin{tabular}{|l|c|c|}  \hline\hline
                                          &$\lambda_1$    &$\lambda_2$      \\ \hline\hline
LSUB4 &1.312 &2.241   \\ \hline
LSUB5 &1.370 &2.030   \\ \hline
LSUB6 &1.357 &2.185   \\ \hline
LSUB7 &1.375 &2.105   \\ \hline
LSUB8 &1.370 &2.145   \\ \hline\hline
SWT	 \cite{chub}      &1.248	                 &2.145                    \\ \hline
Exact Diagonalisations \cite{squareTriangleED} &1.38                    &2.16                      \\ \hline
\end{tabular}
\end{center}
%}
\label{tab1}
\end{table}

\begin{figure}
\centerline{\epsfxsize=7.5cm\epsffile{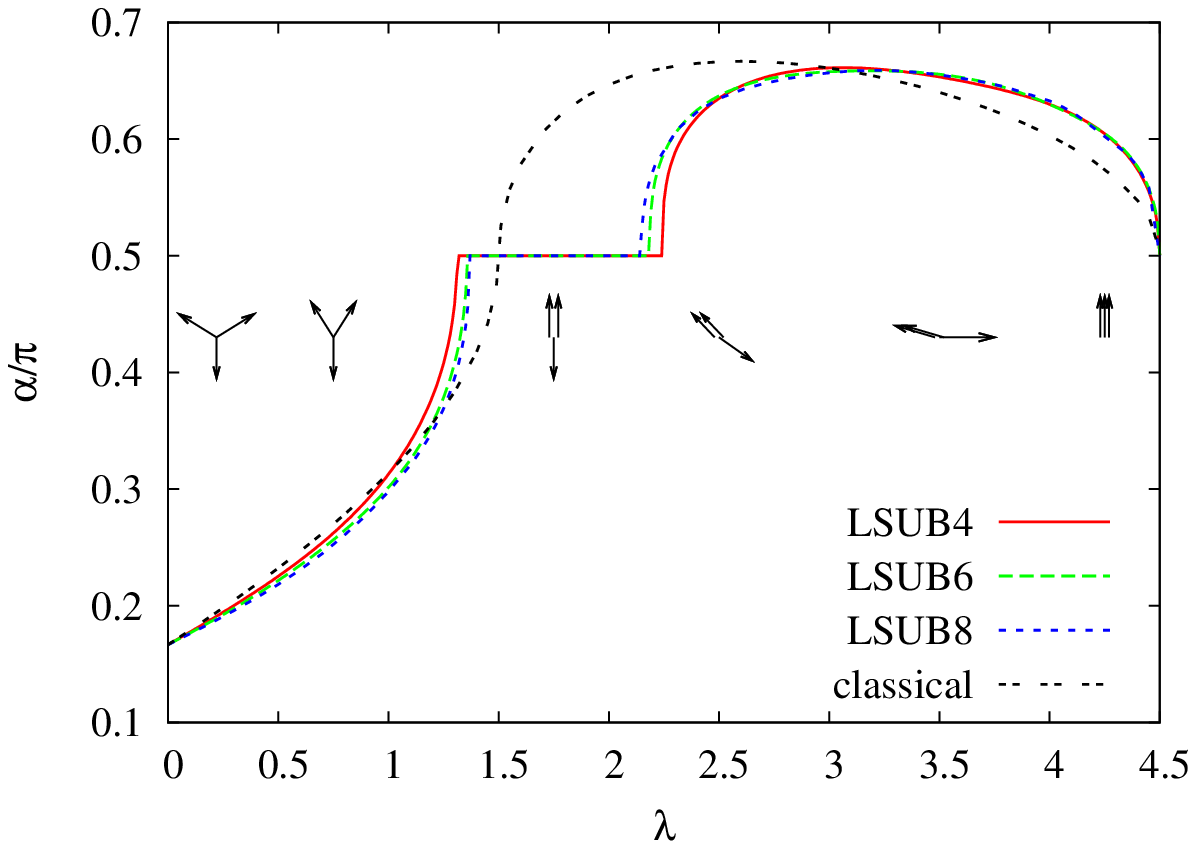} \hspace*{0.5cm}
            \epsfxsize=7.5cm\epsffile{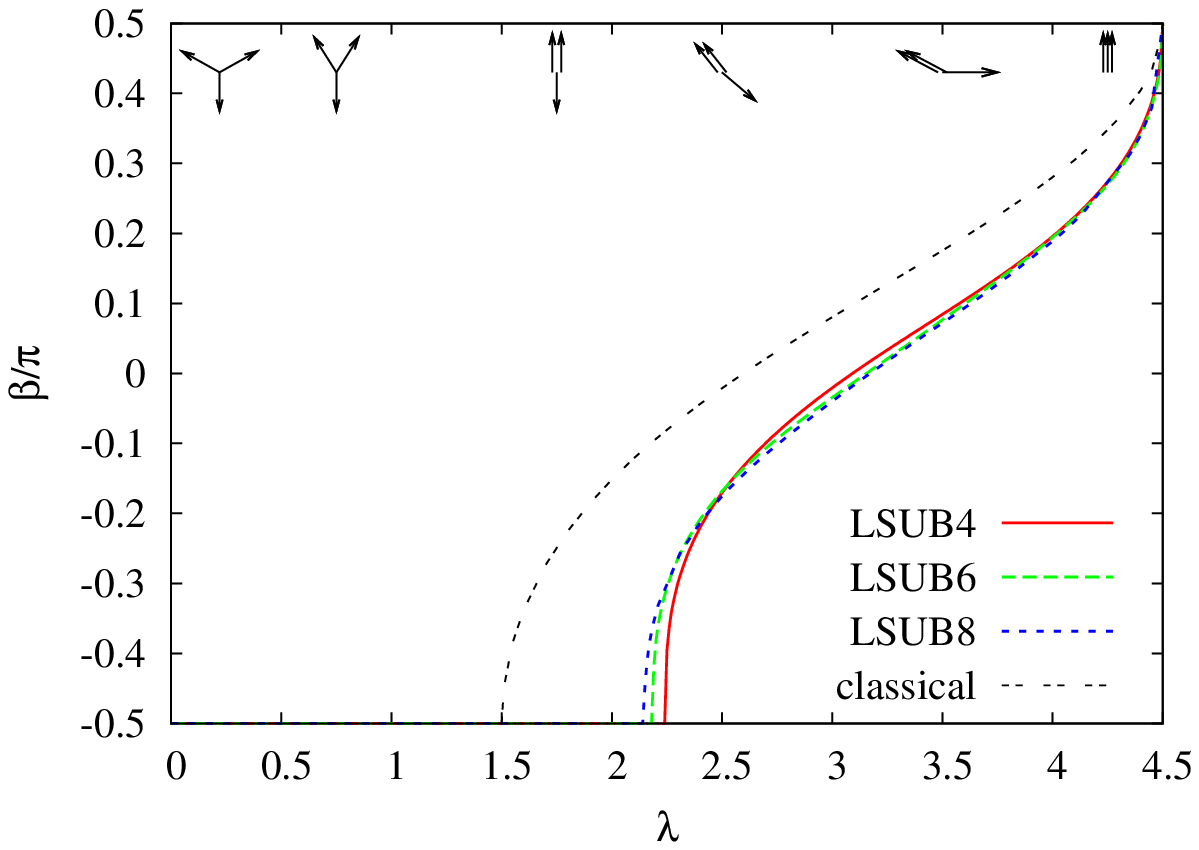}}
\caption{Results for the angle $\alpha/\pi$ (left) and $\beta/\pi$ (right) 
in the model state for the spin-half triangular-lattice 
Heisenberg antiferromagnet in the presence of an external 
magnetic field of strength $\lambda$. The arrows illustrate the actual spin directions.}
\label{triangle_alpha_beta}
\end{figure}

%We remark again that the existence of other plateaux at higher values of 
%$M/M_s$ equal to 1/2, 5/9 and 2/3 have been postulated by recent experiments 
%\cite{fortune} for the magnetic material Cs$_2$CuBr$_4$. The authors of this 
%article suggest that this might be due to unit cells of differing size for the 
%different plateaux, each cell having an overall magnetisation of 1/2. 
%The application of the CCM could be amended straightforwardly to
%accommodate larger  unit cells and so to treat such higher plateaux states,
%although the practical application of the CCM is therefore also more 
%involved. Also, this might also necessitate the minimisation of the CCM
%ground-state energy with respect to more than two parameters in 
%the Hamiltonian. For both of these reasons, the higher plateaux would 
%present more complicated calculations than those presented here. The 
%consideration of the higher plateaux shall form the contents of another 
%article.

We discuss next the canting angles $\alpha$ and $\beta$ 
in the model states I, II, III (see Fig.~\ref{model_states}(a), (b),
(c)) shown 
in Fig.~\ref{triangle_alpha_beta}. 
Note again that to the best of our knowledge data for the angles  have not been presented 
previously by other authors.
A strong difference between the results of the classical 
system and those results of the quantum system is again
obvious, in particular, in the plateau region  where in the quantum model
$\alpha$ and $\beta$ are constant but both angles  change rapidly for the classical
model.
We see that the results for both $\alpha$ and $\beta$ vary continuously, 
although not smoothly, for all values of $\lambda$. There is no sudden discontinuity 
in the solution for the angles as was reported, e.g., for spiral phases
of some frustrated quantum spin models.
% \cite{ccm17,ccm25,ccm34}.
Note that above the plateau the angle $\alpha$ does not vary monotonously
with field. Rather it first increases to $\alpha > \pi/2$ reaching at
maximum at about $\lambda \sim 3.2$. Approaching the saturation then $\alpha$ rapidly decreases
to $\alpha = \pi/2$.  As far as we aware, no such equivalent experimental
results exist for the sublattice magnetizations or tilting angles.
We recommend therefore that experimental investigations of these aspects
of the magnetisation with external field also be carried out.

\begin{figure}
\epsfxsize=11cm
\centerline{\epsffile{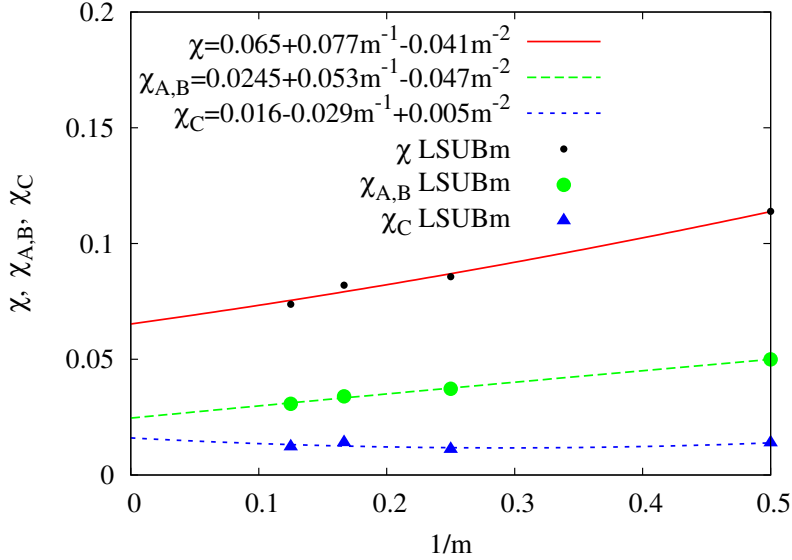}}
\caption{LSUB$m$ results for the 
zero-field uniform susceptibilities $\chi({\lambda \rightarrow 0})$
for the spin-half triangular-lattice  Heisenberg
antiferromagnet with $m=\{2,4,6,8\}$ and
the polynomial fit according to $\chi(m)=c_0+c_1/m+c_2/m^2$.}
\label{extrap_triangle_chi}
\end{figure}

For the zero zero-field uniform susceptibility $\chi({\lambda \rightarrow 0})$, see
Eq.~(\ref{susc}), we obtain 
 $\chi$=$0.1139$, $0.08568$, $0.08200$,
and $0.07378$
for the LSUB$m$ approximation with $m=2,4,6$, and $8$.
In addition, we can also calculate the individual response of 
the sublattices on the magnetic field, i.e.
$\chi_{A,B,C}=\frac{1}{6}\frac{dM_{A,B,C}}{d\lambda}$. Due to the relation
$M= (M_A+M_B+M_C)/3$ we have $\chi= \chi_A+\chi_B+\chi_C$. 
Again we can extrapolate the data for the susceptibilities to $m \to \infty$ using 
$\chi(m)=c_0+c_1/m+c_2/m^2$. The corresponding extrapolation
then yields $\chi=0.065(23)$. (The number in 
brackets indicate the standard deviation.) 
We see from Fig.~\ref{extrap_triangle_chi} that this procedure is 
a reasonable method of extrapolation of the data for the triangular lattice, 
although it is not as good as for the square lattice. This is demonstrated 
by the magnitudes of the estimated standard deviations for the
extrapolated values of $\chi$ for the square and triangular lattices
(of order approximately $10^{-3}$ and $10^{-2}$, respectively).  
We see from Fig.~\ref{extrap_triangle_chi}
that the main contribution to $\chi$ comes from the sublattices $A$ and $B$.
That is not surprising, since for the model state I, see
Fig.~\ref{model_states}b,  
the direction of the magnetisation on the sublattice $C$ is fixed, whereas
the spins on sublattices $A$ and $B$ are rotated towards the field
direction. Indeed, we find that $\chi_{A,B}=0.0245(54)$ and $\chi_C=0.016(13)$
by extrapolating the susceptibilities on the different sublattices separately
(see Fig.~\ref{extrap_triangle_chi}). This analysis leads again to an 
overall value for $\chi(=\chi_A+\chi_B+\chi_C)$ of $\chi=0.065$.         
We can compare this result with $\chi=0.0794$ obtained with 
spin-wave theory \cite{chub,chub94}. (We remark that this value of $\chi$ in 
Ref. \cite{chub94} was referred to as $\chi_\perp$  in this article and furthermore 
that it was defined per volume.)  
Although the magnitudes of $\chi$ for the extrapolated CCM value and the spin-wave 
result agree, the difference between them is still obviously quite large. 
We believe that this difference might be attributed to a somewhat less 
reliable extrapolation (shown clearly in Fig.~\ref{extrap_triangle_chi}) than 
that presented for the square lattice above. However, we should note also that 
the spin-wave theory calculations of Ref.~\cite{chub94} were only ever carried out to order $1/S$. 
(By contrast, the spin-wave theory calculations for the square lattice were carried out 
 to order $1/S^2$ \cite{hamer}.) Hence, both higher order spin-wave results as 
well as higher order CCM-LSUB$m$ results are recommended in order to establish 
a more accurate figure for $\chi$ for the triangular-lattice case and, thus, to resolve
this difference. 

Again we mention that the zero-field uniform susceptibility $\chi({\lambda \rightarrow 0})$,
together with the ground state energy, the sublattice 
magnetisation, the spin stiffness, and the spin-wave velocity constitute the 
fundamental parameter set that determines the low-energy physics of
magnetic systems. Corresponding CCM results for the ground state energy, the 
sublattice magnetisation, the spin stiffness for the triangular-lattice Heisenberg
antiferromagnet at $\lambda=0$ can be found in 
Refs. \cite{ccm12,ccm27}.

\begin{figure}
\epsfxsize=11cm
\centerline{\epsffile{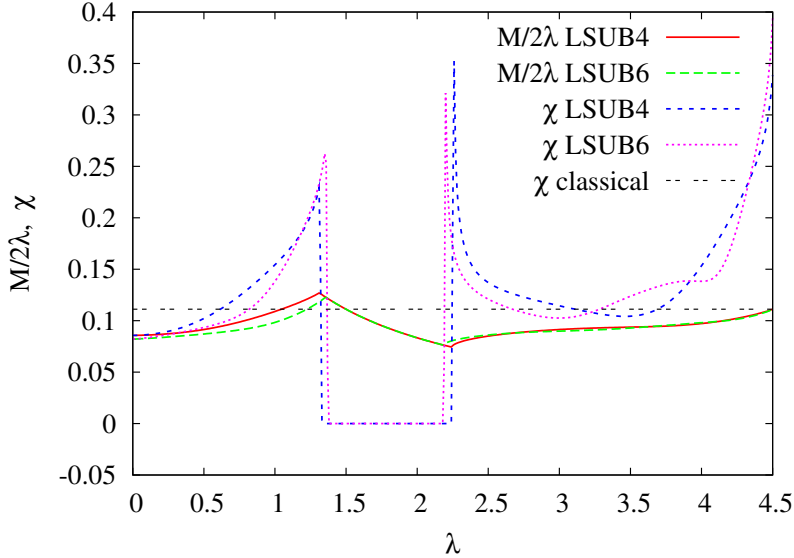}}
\caption{Susceptibility $\chi$, see Eq.
(\ref{susc}), and  the quotient $M/2\lambda$ in dependence on the magnetic
field $\lambda$ the for the spin-half triangular-lattice  Heisenberg
antiferromagnet.}
\label{chi_vs_H_trian}
\end{figure}

As for the square-lattice case above, 
we also present results at the LSUB4 and LSUB6 levels of approximation 
for the field dependence 
of $\chi$  in Fig.~\ref{chi_vs_H_trian}. (Note that we have LSUB8 data for $\chi$ 
only for small fields due to the enormous computational effort of carrying out 
this calculation.)  
Again we compare $\chi(\lambda)$ with $M/(2\lambda)$ which is often 
determined  in experiments and also with the 
classical value $\chi_{\rm clas}=1/9$  that is independent of
$\lambda$.

From Fig.~\ref{chi_vs_H_trian} it is obvious that
$\chi$ and $M/2\lambda$ agree well with each other up to about
$\lambda=\lambda_s/10$
(the difference is about 7\% at $\lambda=0.45$), but deviate significantly
for larger $\lambda$.
As for the square lattice  $\chi$  grows with $\lambda$ starting from zero
field 
up to the bottom of the plateau at $\lambda_1$. In the plateau region $\chi$ is zero indicating a finite
excitation gap about the plateau ground state. Approaching the plateau from
below or from above $\chi(\lambda)$ exhibits a sharp peak.
Such peaks at the
end of the plateau are indeed observed in experiments on an antiferromagnet
on the triangular lattice,  see e.g. Figs.~9 and 10 in Ref.
\cite{ono}.
Between the top of the plateau at $\lambda_2$ and the saturation at
$\lambda_s$ we find a broad region where  the
susceptibility is small $\chi \approx 0.1$. Approaching the saturation
$\chi$
again becomes large.
The oscillations seen for $\lambda \sim 3.5$
seem to be an artefact of CCM-LSUB$m$ approximation. 
However, we expect again that the amplitude of oscillation will decrease 
with increasing approximation level and would disappear entirely 
in the limit $m \rightarrow \infty$.

\section{Conclusions}

In this article we describe how the coupled cluster method (CCM) 
may be applied in order to calculate the behaviour of quantum 
antiferromagnetic systems in the presence of external magnetic 
fields. We have determined 
the ground-state energy, the total lattice magnetisation as well as
sublattice magnetisations and the uniform susceptibility for the 
spin-half Heisenberg antiferromagnets  on the square lattice and 
the triangular lattice  by using the CCM to high orders of approximation. 
We showed that high-order CCM calculations give reasonable results for 
these quantities over all values of the magnetic field strength 
$\lambda$ for both lattices. For example, the CCM result for the lattice 
magnetisation for the square lattice compare well to QMC and spin-wave 
theory results for all values of the magnetic field strength. Our 
result for the uniform susceptibility of $\chi=0.070$ for the square lattice 
is in reasonable agreement with those results of other methods (e.g., 
$\chi=0.0669(7)$ via QMC). Again, we believe that even closer agreement 
would occur with high orders of LSUB$m$ approximation.

CCM results presented here for the total 
lattice magnetisation for the triangular lattice show the characteristic 
magnetisation plateau at $M/M_s=\frac 13$ also seen in other 
studies \cite{Hon1999,squareTriangleED,chub,HSR04}. The width of this plateau was 
estimated by us to be given by $1.37 \lesssim \lambda \lesssim 2.15$.
 This result was found to be in good agreement with 
results of spin-wave theory  \cite{chub} ($1.248 < \lambda <  2.145$) and 
exact diagonalisations \cite{Hon1999,HSR04,nishi,squareTriangleED} ($1.38 < \lambda < 2.16$). 
Our results therefore support those of exact diagonalisations
that indicate that the plateau begins at a higher value
of $\lambda$ than that suggested by spin-wave theory. 
In addition, we provide results for sublattice magnetisations 
$M_A$, $M_B$, and $M_C$ evaluated on the individual sublattices 
$A$, $B$, and $C$ of the triangular lattice that allows a better 
understanding of the magnetisation process of the triangular lattice.
As far as we are aware, this is the first time that results
for the individual sublattice magnetisations (and angles)
have been presented. 
Our result for the longitudinal uniform low-field susceptibility 
$\chi=0.065$ compares to the result of 
result of spin-wave theory ($\chi=0.0794$), i.e. there is quite  
a large difference between the spin-wave and the CCM result.  
 Hence, higher order approximations for both SWT and CCM LSUB$m$ 
 calculations and/or alternative approaches 
are recommended in order to obtain more reliable values  for 
$\chi$ for the triangular-lattice case.
The susceptibility $\chi(\lambda)$ in dependence on the magnetic field
$\lambda$ shows for the triangular lattice
characteristic
sharp peaks at the bottom and the top of the plateau which may be used as
indicators in experiments for a magnetisation plateau.\\

{\bf Acknowledgements: } 
The present study was supported by the DFG (project Ri615/18-1).
We are indebted to the research group of S. Mertens for providing us access to
their Tina - Beowulf-Cluster-Computer. 
DJJF gratefully acknowledges support for the research presented 
here from the European Science Foundation 
(Research Network Programme: Highly Frustrated Magnetism).

\pagebreak

\pagebreak


\begin{thebibliography}{200}
\bibitem{defin_chi} $\chi$ is sometimes defined per volume, see e.g.
Ref.~\cite{chub94}. Although this factor is unity for the square lattice, it
yields a different factor of $2/\sqrt{3}$ for the triangular lattice.
\end{thebibliography}
\end{document}